\documentclass[aps,prb,twocolumn,showpacs]{revtex4}
\usepackage{bm}
\usepackage{epsfig}
\usepackage{amsmath}

\newcommand{\bpm}{\begin{pmatrix}}
\newcommand{\epm}{\end{pmatrix}}
\newcommand{\ba}{\begin{eqnarray}}
\newcommand{\ea}{\end{eqnarray}}

\begin{document}

\title{Evidence for a charge collective mode associated with superconductivity in copper oxides from neutron and x-ray scattering measurements of La$_{2-x}$Sr$_x$CuO$_4$}

\author{S. R. Park$^{1,2}$, T. Fukuda$^{3,4}$, A. Hamann$^5$, D. Lamago$^{5,6}$, L. Pintschovius$^5$, M. Fujita$^7$, K. Yamada$^8$, D. Reznik$^{1}$}

\email[Electronic address:$~~$]{dmitry.reznik@colorado.edu}

\affiliation{$^1$Department of physics, University of Colorado at
Boulder, Boulder, Colorado 80309, USA}

\affiliation{$^{2}$Department of Physics, Incheon National University, Incheon 406-772, Korea}

\affiliation{$^3$Materials Dynamics Laboratory, RIKEN SPring-8 Center, Sayo, Hyogo 679-5148, Japan}

\affiliation{$^4$Synchrotron Radiation Research Unit, SPring-8/JAEA, Hyogo 679-5148, Japan}

\affiliation{$^5$Karlsruhe Institute f\"ur Technologie, Institute f\"ur Festk\"orperphysik, P.O. Box 3640, D-76021 Karlsruhe, Germany}

\affiliation{$^6$Laboratoire Leon Brillouin, CEA-Saclay, F-91191 Gif sur Yvette Cedex, France}

\affiliation{$^7$Institute for Materials Research, Tohoku University, Senda, Miyagi 980-8577, Japan}

\affiliation{$^8$Institute of Materials Structure Science, High Energy Accelerator Research Organization (KEK), Tsukuba, Ibaraki 305-0801, Japan}

\date{\today}

\begin{abstract}
In superconducting copper oxides some Cu-O bond-stretching phonons around 70meV show anomalous giant softening and broadening of electronic origin and electronic dispersions have large renormalization kinks near the same energy. These observations suggest that phonon broadening originates from quasiparticle excitations across the Fermi surface and the electronic dispersion kinks originate from coupling to anomalous phonons. We measured the phonon anomaly in underdoped (x=0.05) and overdoped (x=0.20,0.25) La$_{2-x}$Sr$_x$CuO$_4$ by inelastic neutron and x-ray scattering with high resolution. Combining these and previously published data, we found that doping-dependence of the magnitude of the giant phonon anomaly is very different from that of the ARPES kink, i.e. the two phenomena are not connected. We show that these results provide indirect evidence that the phonon anomaly originates from novel collective charge excitations as opposed to interactions with electron-hole pairs. Their amplitude follows the superconducting dome so these charge modes may be important for superconductivity.
    
\end{abstract}
 \maketitle


Lattice vibrations in metals can be damped and/or softened by either electronic quasiparticles or collective charge excitations (e.g. plasmons). Giant phonon softening and line broadening of electronic origin of the longitudinal Cu-O bond stretching phonons near half-way to the zone boundary (giant anomaly) was observed in copper oxide high temperature superconductors (HTSCs) \cite{Dmitry_Review,Dmitry_Review2,McQueeney1999,Pintschovius1999, Fukuda, Reznik2006, Pintschovius2006, Pintschovius2004,Reznik2008, Uchiyama, Graf}. It was previously found at superconducting compositions and was also absent in undoped and overdoped nonsuperconducting copper oxides \cite{Fukuda, Reznik2006,Pintschovius2006}. 

First reports interpreted the phonon anomaly as a signature of unit cell doubling \cite{McQueeney1999} followed by a different interpretation \cite{Reznik2006} in terms of coupling of the phonon to dynamic charge stripes. Subsequently, close kinematic relationship between the longitudinal Cu-O bond stretching mode dispersion and renormalization of electronic quasiparticles in Bi$_2$Sr$_{1.6}$La$_{0.4}$CuO$_6$ (Bi2201) indicated that the phonon anomaly may originate from the coupling of phonons to electronic quasiparticles \cite{Graf}. $t-J$ model-based calculations also predicted strong coupling of optical phonons to electron-hole excitations \cite{Nagaosa1}. 

If phonons couple strongly to electronic quasipaticles, then a BCS-type mechanism of superconductivity may still be valid.  Alternatively, the phonon anomaly would arise from a novel collective charge excitations at low energies (at least 70 meV). Then such collective mode may  provide the pairing interaction \cite{Alexander, Monthoux2004, Monthoux2007}. 

We measured Cu-O bond-stretching phonons in La$_{2-x}$Sr$_x$CuO$_4$ for undoped x=0.00, nonsuperconducting underdoped x=0.05, and superconducting overdoped x=0.20, 0.25.  x=0.05 did not show characteristic signatures of the giant phonon anomaly.  x=0.20 showed a very strong phonon anomaly, which was dramatically reduced already at x=0.25. On the other hand, the magnitude of the kink in the electronic dispersions gradually decreases from x=0.05 to x=0.30 \cite{Park2,Zhou1}. The new data, combined with previously published data for other x, allowed us to correlate the doping dependence of the giant phonon anomaly with that of the the electronic dispersion kinks and of the superconducting transition temperature T$_c$. We show that the amplitude of the giant phonon anomaly tracks T$_c$ but not the strength of the ARPES kink. We conclude that the phonon anomaly does not originate from enhanced coupling to quasiparticles. Instead our results indirectly indicate that anomalous phonons may be interacting with a novel charge collective mode that may be relevant to the superconductivity mechanism.

\begin{figure*}
\centering \epsfxsize=18.2cm \epsfbox{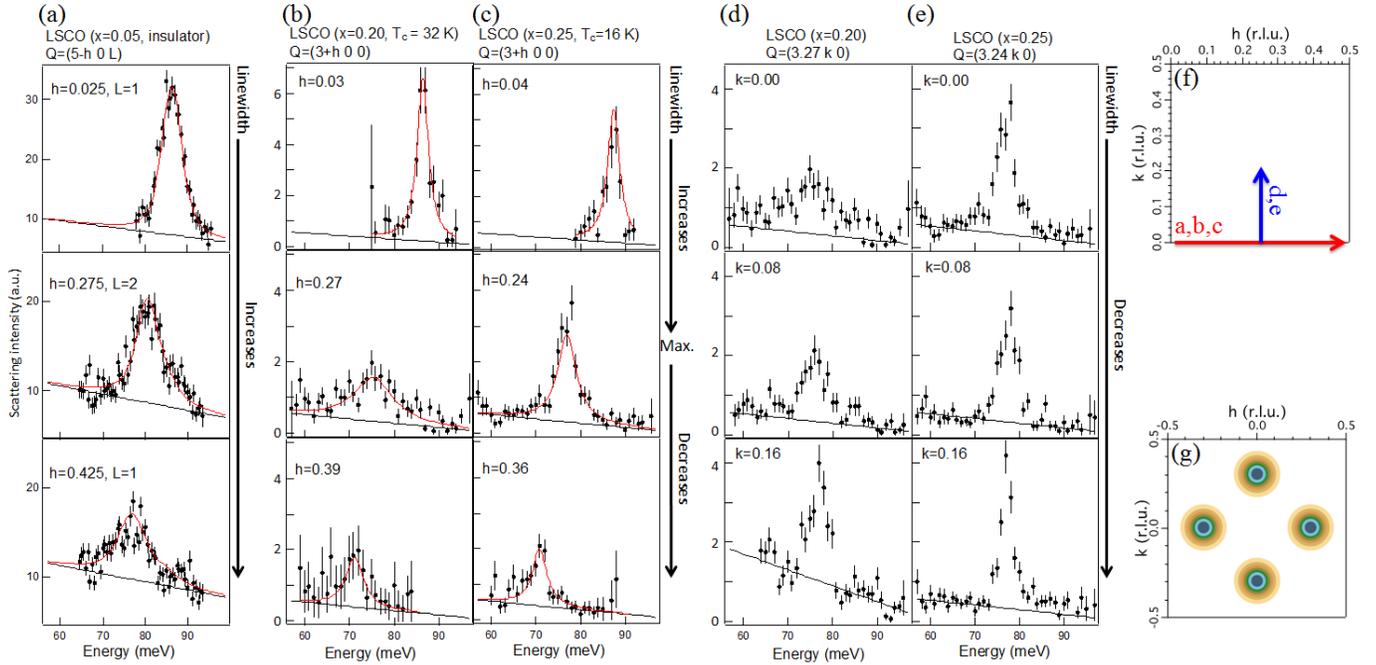} \caption{(a) INS data of La$_{2-x}$Sr$_x$CuO$_4$ (x=0.05) at Q=(5-h, 0, 1) (top and bottom panels) and at Q=(5-h, 0, 2) (middle panel). L-values were chosen to avoid contaminations. See Ref. [\onlinecite{Park1}] for details of the fits represented by solid line. (b,c) IXS data of La$_{2-x}$Sr$_x$CuO$_4$, (b) x=0.20 and (c) x=0.25 at Q=(3+h, 0, 0). Red lines represent the Lorentzian fitting function. (d,e) Phonon dispersion in LSCO from Q=(3.27, 0, 0) (d) and Q=(3.24, 0, 0) (e) in the [0 1 0]- direction, for (d) x=0.20 and (e) x=0.25. (f) Schematic of the dispersion directions in (a)-(e). (g) Qualitative picture of anomalous phonon broadening distribution.}\label{fig2}
\end{figure*}

Inelastic neutron-scattering (INS) experiments were performed on large high-quality single crystals of LSCO (x=0.05) on the 1-T triple-axis spectrometer at the ORPHEE reactor at the Laboratoire Leon Brillouin at Saclay, France. The monochromator/analyzer were Cu220/PG002 respectively. The measurements were performed at reciprocal lattice vectors Q = (5 - h 0 L), L=0,1,2, in the tetragonal notation. Inelastic X-ray scattering (IXS) experiments on LSCO (x=0.00, x=0.20 and x=0.25) were performed at BL35XU at SPring-8 \cite{Baron00} in the same experimental conditions as in a previous study \cite{Reznik2008_2}. For all measurements temperature was near 10K. Neutron data for LSCO (x=0.07, 0.15, 0.30) as well as the ARPES results were published previously \cite{Reznik2006,Pintschovius2006,Park2}. 

The phonons disperse downward from the zone center (h=0) to the zone boundary (h=0.5) for all three samples in the [1 0 0]-direction indicated by the red arrow in Fig. 1(f). For the strongly underdoped insulating x=0.05 sample \cite{Takagi}, the linewidth monotonically increases from the zone center to the zone boundary (Fig. 1a). In overdoped x=0.20 and x=0.25, the phonons broaden from h=0.03 to h=0.27 and then sharpen towards h=0.39 (Fig. 1b,c).  

Giant linewidth broadening rapidly decreases at x=0.20 away from k=0 (Fig. 1(d,e) ). Even though the phonon at x=0.25 shows a much weaker anomaly, peak sharpening at k=0.16 is still clear compared to k=0 (Fig. 1e). Therefore, the giant phonon anomaly in copper oxides is concentrated near reduced wavevectors q=(h,k,l) for h=0.25,0.3 and k=0. Previous work showed that it is independent of l.

Increasing linewidth towards the zone boundary at x=0.05 is consistent with inhomogeneous doping as discussed below \cite{Park1}. Linewidth maximum half way to the zone boundary observed in x=0.20 and x=0.25 cannot be explained by this \cite{supp} or any other simple mechanism \cite{Dmitry_Review, Dmitry_Review2}. 

\begin{figure}
\centering \epsfxsize=8.7cm \epsfbox{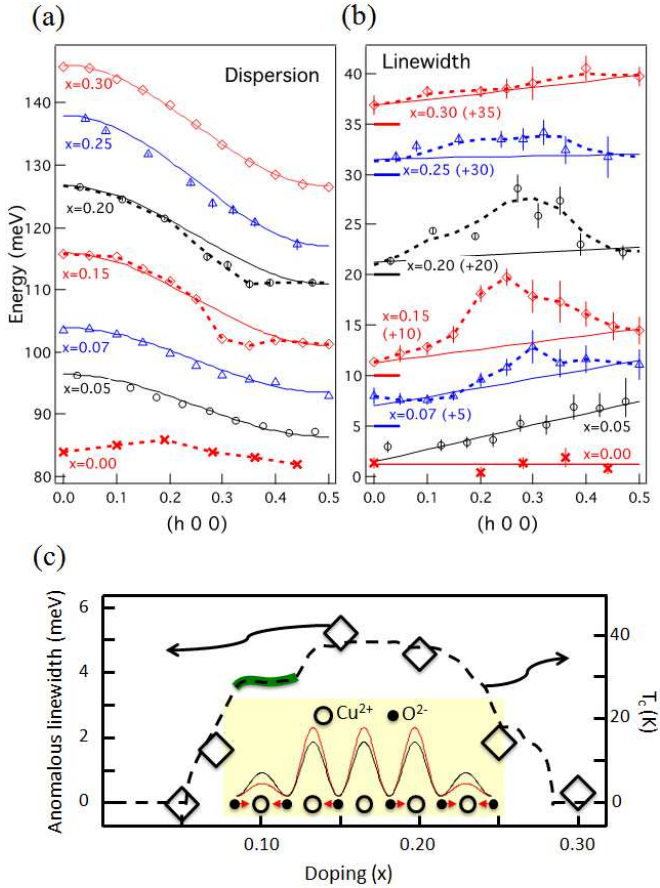} \caption{(a) Longitudinal Cu-O bond stretching phonon dispersions. Solid lines indicate cosine functions. Dashed lines are guides to eye. The data for x=0.05, 0.07, 0.15, 0.20, 0.25, and 0.30 are shifted by 10, 20, 30, 40, 50 and 60 meV for clarity, respectively. (b) Resolution-corrected linewidths of the longitudinal Cu-O bond stretching mode. Solid straight lines indicate "background" contributions to linewidths that smoothly increase towards the zone boundary as described in the text. The data for x=0.07, 0.15, 0.20, 0.25, and 0.30 are shifted by 5, 10, 20, 30, and 35 meV, respectively. (c) Average linewidths of h=0.20, 0.25, 0.30 and 0.35 (diamonds) after subtracting background linewidths (solid lines in Fig. 2(b)) plotted as a function of doping. Dashed line represents $T_c$ of LSCO \cite{Ikeuchi}. Green solid line indicates a plateau of $T_c$ that may originate from competition with another order parameter such as stripes. Inset illustrates strong coupling between Cu-O bond stretching mode and electronic density fluctuations at h=0.25. Black and red lines indicate charge densities with and without atomic displacements (red arrows), respectively. Red arrows indicate atomic displacements of the anomalous phonon.}\label{fig2}
\end{figure}

In order to isolate the giant phonon anomaly from the overall broadening towards the zone boundary, we drew straight lines connecting the linewidth data at the zone center and the zone boundary (Fig 2b). Their upward slope decreases from lower to higher doping. Linewidths of x=0.05 and x=0.30, which are nonsuperconducting, are very close to the straight lines, whereas the other dopings clearly deviate above the lines with the maximum deviation close to h=0.25 or h=0.3. Giant phonon anomaly gets its name from the huge effect for 0.1$\leq$x$\leq$0.2.  Softening below the sinusoidal dispersion was also observed in superconducting samples (Fig. 2a). This softening approximately scales with the broadening, which is qualitatively consistent with the Kramers-Kronig relation between real and imaginary parts of the phonon self-energy. The caveat here is that the phonon dispersion in the absence of the phonon anomaly is not precisely known. DFT calculations for optimal doping show sinusoidal dispersion similar to what is actually observed at x=0.30 \cite{Dmitry_Review}, On the other hand, the experimental dispersion in undoped La2CuO4 where both the giant phonon anomaly and the inhomogeneous doping effect are absent, clearly deviates from the sine function (Fig. 2a).

The overall increase of the phonon linewidth to the zone boundary becomes more pronounced with underdoping (Fig. 2b) with the biggest increase at x=0.05 as well as at x=0.04 in the data of Fukuda et al. \cite{Fukuda}, which was unexplained. It was shown in Ref. \cite{Park1} that this effect originates from inhomogeneous doping combined with doping-induced phonon softening at Sr concentrations below optimal doping. Since the doping-dependence of the phonon frequency increases towards the zone boundary, the width of the measured phonon spectrum increases smoothly towards the zone boundary as well. We were able to explain all the data at x=0.05 using this model. However, at x=0.07, which is already superconducting, there is an additional linewidth around h=0.3 that cannot be explained by inhomogeneous doping effect \cite{supp}. This extra broadening becomes very large around optimal doping.

We can phenomenologically isolate the giant phonon anomaly from other contributions to the phonon linewidth by subtracting the "background" linewidth indicated by the solid lines in Fig. 2(b). Fig. 2(c) shows the average phonon linewidth at h=0.20, 0.25, 0.30 and 0.35 after subtracting this background linewidth as a function of doping. Remarkably, the linewidths of the giant phonon anomaly at low temperatures follow the superconducting transition temperature ($T_c$). 

\begin{figure}
\centering \epsfxsize=8.7cm \epsfbox{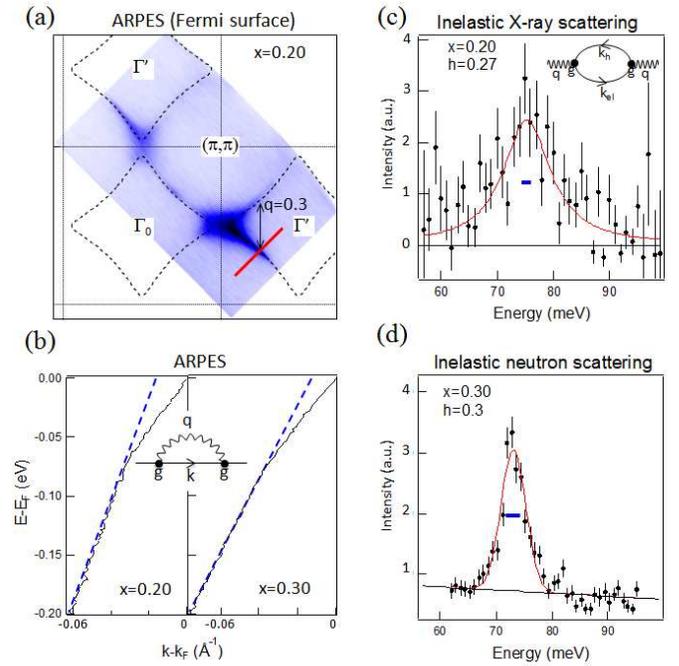} \caption{(a) Fermi surface (FS) of LSCO, x=0.20, measured by ARPES. Double sided arrows indicate the wavevector (h=0.3) of the giant phonon anomaly. This q-vector connects only two points of FS, indicating no FS nesting. (b) Electronic quasiparticle dispersions of LSCO (x=0.20) and (x=0.30) along the red line in (a), where Fermi momentum is connected to the Fermi momentum on the other side of the arrow by h=0.3. Dashed straight lines represent band velocities from -0.2 to -0.1 eV for x=0.20 and x=0.30, respectively. Inset is the Feynman diagram for electronic quasiparticle propagation with electron-phonon coupling. (c) Background-subtracted IXS data of x=0.20. Inset is the Feynman diagram for phonon propagation with electron-phonon coupling. (d) INS data of x=0.30. Blue horizontal bars inside peaks in (c,d) show experimental resolutions. Vertex, g, in the Feynman diagrams in (b) and (c) is the same.}\label{fig2}
\end{figure}

If strong interactions between electronic quasiparticles and phonons are responsible for the giant phonon anomaly, they should also renormalize electronic dispersions. A kink in the electronic dispersions around 70 meV has been reported previously by ARPES and the possibility that the giant phonon anomaly and ARPES features are two sides of the same coin has been raised \cite{Graf, Reznik2008_3}. 

In order to see if electronic quasiparticles are responsible for the giant phonon anomaly, we first checked if the phonon wavevector spans parallel sheets of the Fermi surface, near h=0.25 or 0.3, since such FS nesting can give the phonon anomaly \cite{Renker}. Fig. 3(a) illustrates that there is no nesting near h=0.25 or h=0.3. This result contrasts the situation in BSCCO where the FS is turned by 90 degrees and FS nesting is possible \cite{Graf}.

Strong q-dependence of the electron-phonon coupling matrix element can induce phonon anomalies in materials without FS nesting \cite{Heid, Lamago1, Weber2}. The same matrix element should enhance the kink in dispersions of electronic quasiparticles coupled to the phonon. The (red) solid line in Fig. 3(a) crosses $k_F$ spanned by the wavevector of the giant phonon anomaly. Fig. 3(b) shows that the kink strength of x=0.30 is reduced by only about 30 $\%$ from that in x=0.20 \cite{Park2}. On the contrary, the phonon broadening is  strongly suppressed already by x=0.25 (Figs. 1,2) and completely disappears at x=0.30 \cite{Pintschovius2006} (Fig. 2(b)). This is clear evidence that neither the giant phonon anomaly nor the ARPES kink originate from the interaction between the anomalous phonons and electronic quasiparticles. Magnetic fluctuations observed by neutron scattering also cannot be responsible for the 70 meV kink, because they are not present below 90meV for x=0.3\cite{Wakimoto}. Previous work also ruled out significant anharmonic contribution to the anomaly \cite{Dmitry_Review}. 

Note that we cannot rule out a different phononic mechanism behind the ARPES kink: although inhomogeneous doping effect should become negligible above optimal doping, some broadening towards the zone boundary remains in optimal and overdoped samples. This extra broadening is smaller than the inhomogeneous doping effect discussed above, and is much smaller than the giant phonon anomaly, but it is still significantly larger than the linewidths calculated by LDA \cite{Reznik2008_3}. The origin of this extra broadening is not completely understood, but it can be due to the coupling to electronic quasiparticles.

Here we propose that the giant phonon anomaly instead originates from collective charge excitations, because present work rules out other reasonable possibilities. Up to this point there is no clear and direct evidence for such fluctuations at low energies although there is other indirect and circumstantial evidence. For example, low-energy collective charge excitations can exist in LSCO as fluctuating stripes \cite{KivelsonReview}, even though they have not been directly observed yet \cite{note1}. In fact the giant phonon anomaly is also very strong in compounds with static stripes \cite{Tranquada1995, Abbamonte,Kivelson1990}, La$_{1.6-x}$Nd$_{0.4}$Sr$_x$CuO$_4$ \cite{Reznik2008_2} and La$_{1.875}$Ba$_{0.125}$CuO$_4$ (LBCO) \cite{Reznik2006}. Plateau of $T_c$ of LSCO near x=0.10 (green line in Fig. 2c) may also be due to superconductivity competing with stripes \cite{Koike}. Based on static stripes, low energy collective charge excitations have been predicted to exist in copper oxides  \cite{Kaneshita,Mukhin,Bulut}. Recent inelastic neutron scattering experiments on stripe-ordered nickelates revealed low-energy collective charge excitations, which also vertically disperse from the charge stripe ordering wave vector \cite{Sveta}. Even if static stripes do not exist in LSCO, it is reasonable to suppose that they survive as fluctuations. Alternative low energy collective charge modes may also be possible. Recent resonant X-ray scattering studies on YBCO also found low energy charge fluctuations interpreted as CDW fluctuations  \cite{Blackburn, Keimer} close to the wave vector of the giant phonon anomaly. Results presented here demonstrate that these collective charge excitations may underlie the superconductivity mechanism, because their signature in the phonon spectra scales with T$_c$.

These collective charge excitations may be universal in copper oxide superconductors, since the giant phonon anomaly has been also observed in YBa$_2$Cu$_3$O$_{7-\delta}$ (YBCO), HgBa$_2$CuO$_{4+\delta}$ (HBCO), Bi2201, and Ca$_{2-x}$CuO$_2$Cl$_2$ \cite{Pintschovius2004, Reznik2008, Uchiyama, Graf, d'Astuto13}.

Our experiments revealed indirect evidence for collective charge excitations in copper oxides whose amplitude correlates with the superconducting dome.  They  may be crucial for superconductivity in LSCO, since the giant phonon anomaly is closely related to $T_c$. It has been theoretically proposed that collective charge excitations can meditate Cooper pairs in heavy fermions and copper oxides \cite{Monthoux2004, Monthoux2007}. Indeed, the collective-charge-excitation mediated superconductivity may be realized in CeCu$_2$Si$_2$ \cite{HQYuan, Alexander}. Such theories should be constrained by our results that the collective charge excitations in copper oxides, within the phonon energy scale, are concentrated near the wave vector of the giant phonon anomaly, h=0.25 or h=0.30. In order to make further progress, it will be necessary to investigate these excitations by high resolution resonant inelastic x-ray scattering, which has not yet been developed. We hope that this result will stimulate further development of this technique.

\acknowledgements
The authors thank A.Q.R. Baron for help with the IXS measurements and comments on the manuscript. The work at Tohoku University was supported by the Grant-In-Aid for Science Research A (22244039) from the MEXT of Japan. S.R.P. and D.R. were supported by the DOE, Office of Basic Energy Sciences, Office of Science, under Contract No. DE-SC0006939. The synchrotron radiation experiments were performed at SPring-8 with the approval of the Japan Synchrotron Radiation Research Institute (JASRI) (Proposal numbers 2007A1441 and 2009A1506 BL No35).

\end{document}